\documentstyle[twocolumn,pra,aps]{revtex}
\def\begmc{}
\def\endmc{}
\def\numeq#1{#1}

\def\Im{\mathop{\rm Im}}
\def\sgn{\mathop{\rm sgn}}

\begin{document}
\title{Quantum fluctuations of vacuum stress tensors and spacetime curvatures}
\author{Marc-Thierry Jaekel $^{a}$ and Serge Reynaud $^{b}$}
\address{(a) Laboratoire de Physique Th\'{e}orique de l'ENS 
\thanks{%
Laboratoire CNRS associ\'e \`a l' ENS et l'Universit\'e Paris-Sud}, \\
24 rue Lhomond F75231 Paris Cedex 05, France\\
(b) Laboratoire Kastler Brossel
\thanks{%
Laboratoire de l' ENS et de l'Universit\'e Pierre et Marie Curie
associ\'e au CNRS}, \\
UPMC case 74, 4 place Jussieu, F75252 Paris Cedex 05, France}
\date{{\sc Annalen der Physik} {\bf 4} (1995) 68-86}
\maketitle

\begin{abstract}
{\bf Abstract} We analyze the quantum fluctuations of vacuum stress tensors
and spacetime curvatures, using the framework of linear response theory
which connects these fluctuations to dissipation mechanisms arising when
stress tensors and spacetime metric are coupled. Vacuum fluctuations of
spacetime curvatures are shown to be a sum of two contributions at lowest
orders; the first one corresponds to vacuum gravitational waves and is
restricted to light-like wavevectors and vanishing Einstein curvature, while
the second one arises from gravity of vacuum stress tensors. From these
fluctuations, we deduce noise spectra for geodesic deviations registered by
probe fields which determine ultimate limits in length or time measurements.
In particular, a relation between noise spectra characterizing spacetime
fluctuations and the number of massless neutrino fields is obtained.

\end{abstract}

\begmc

\section*{Introduction}

Fluctuations of quantum fields lead to observable mechanical effects. Field
quanta carry energy and momentum and exert radiation pressure forces upon
scatterers \cite{Qfvst1}. These forces are themselves fluctuating
quantities, like stress tensors which describe energy and momentum
densities. Such force fluctuations are associated with dissipative forces
which damp the motion of scatterers \cite{Qfvst2}.

Those mechanical effects of quantum fields persist in a state with no
quanta, i.e. in vacuum. Vacuum fluctuations also exert radiation pressure
forces, the so-called Casimir forces, upon scattering boundaries \cite
{Qfvst3}. Such forces fluctuate, like vacuum stress tensors \cite{Qfvst4}.
Relations between fluctuation and dissipation still hold in vacuum, which
may be regarded as the zero temperature limit of a thermal equilibrium state 
\cite{Qfvst5}. Dissipative motional forces in vacuum may be identified \cite
{Qfvst6} with the effect of radiation of energy into vacuum stemming from
non uniform motion of scattering boundaries \cite{Qfvst7}.

When attention is focussed upon questions of principle, it appears that
fluctuation-dissipation mechanisms play a fundamental role for determining
ultimate limits in quantum measurements. This results for example from an
analysis of interferometric length measurements \cite{Qfvst8} or from a
general analysis of the effects of noise and dissipation in high-sensitivity
measurements \cite{Qfvst9}. For measurements performed with endpoints of
mass $m$, radiation pressure exerted by vacuum upon these endpoints imposes 
\cite{Qfvst10} a sensitivity limit of the order of Compton wavelength $\frac
\hbar {mc}$. For macroscopic masses however, i.e. precisely for masses
greater than Planck mass $\sqrt{\frac{\hbar c}{G}}\approx 22\mu $g where $G$
is Newton constant, Compton wavelength is smaller than Planck length  
\[
l_{P}=\sqrt{\frac{\hbar G}{c^{3}}} 
\]
Since it cannot be accepted that sensitivity in spacetime probing goes
beyond Planck length (see for example \cite{Qfvst11}), it appears that
gravity has to be taken into account when analysing ultimate limits.

When it is treated in the same spirit as other field theories \cite
{Qfvst12,Qfvst13}, gravitation exhibits quantum metric and curvature
fluctuations \cite{Qfvst14}. Like classical curvature perturbations
associated with gravitational waves \cite{Qfvst15}, quantum curvature
fluctuations are felt by any field used to probe spacetime. Estimating noise
spectra for fluctuating geodesic deviations stemming from these vacuum
gravitational waves leads to a universal spectrum for length fluctuations
which prevents sensitivity from going beyond Planck length for measurements
performed with macroscopic masses \cite{Qfvst16}.

Einstein equation for gravitation \cite{Qfvst17} can be regarded as a
response equation which describes the metric response to a stress tensor
perturbation, and used to derive vacuum fluctuations of metric. It also
leads to extra curvature fluctuations arising from gravity of quantum
fluctuations of vacuum stress tensors \cite{Qfvst18}. The opinion that
gravity only feels mean values of stress tensors has sometimes been
expressed \cite{Qfvst19}, but it is known to endanger the consistency of
quantum predictions \cite{Qfvst20}. Even if the existence of quantum
fluctuations associated with Einstein equation is denied, unavoidable
coupling to vacuum stress tensors lures metric fluctuations into the quantum
domain.

Fluctuations of vacuum stress tensors are associated with a dissipative
response of vacuum to a metric perturbation that can be identified with
particle production in a curved spacetime \cite{Qfvst21}. It follows that
vacuum stress tensors and spacetime metric have to be regarded as dynamical
systems coupled through dissipative mechanisms, and that the determination
of ultimate quantum limits in spacetime probing requires a consistent
treatment of their fluctuations.

In the present paper, we want to give a consistent description of the
fluctuations of vacuum stress tensors and spacetime curvatures, of the
associated dissipation mechanisms and of the ultimate sensitivity that may
be reached in length or time measurements. For that purpose, we shall rely
on an analogy with the problem of vacuum radiation pressure acting upon
moving scatterers. This analogy has often been used as a guide for studying
the interplay between quantum fluctuations and gravitation \cite{Qfvst22}.
Here, we shall apply techniques of linear response theory, which have
already proved fruitful for studying moving boundaries coupled to vacuum
radiation pressure, to analyse gravitational fluctuations coupled to vacuum
stress tensors.

The paper is organized as follows. In a preliminary section, we recall the
framework of quantum fluctuations and linear response theory and discuss
some questions more specific to gravitational fluctuations. Next sections
are devoted to the study of curvature and stress tensor fluctuations. First,
we derive the curvature fluctuations associated with Einstein equation for
gravitation and recall how fluctuations of vacuum stress tensors may be
computed when metric fluctuations are disregarded. General properties of
these fluctuations based upon Lorentz invariance and conservation laws in
Minkowski spacetime are discussed while explicit results are gathered in
appendix A. We then show how to build a consistent description of coupled
fluctuations of metric and stress tensors, which is similar to that used for
position and force fluctuations for a mirror in vacuum \cite{Qfvst10} and
reflects thermalization of a system coupled to a bath at zero temperature.
To this aim, we assume that the low frequency behaviour of gravitation is
effectively described by Einstein equation; the possibility that vacuum
polarization may modify the effective behaviour of gravitation at low
frequency is discussed separately in appendix B. In final sections, we
deduce noise spectra for the geodesic deviations registered by a probe
field, provide ultimate sensitivity limits that can be attained in length or
time measurements, and discuss the obtained results.

\section*{Quantum fluctuations and linear response theory}

In this preliminary section, we recall general definitions for the
correlation and susceptibility functions associated with quantum
fluctuations, and briefly discuss some questions which arise when linear
response theory is applied to coupled metric and stress tensors.

Correlation functions are defined according to the general prescription  
\begin{eqnarray*}
C_{AB}(x) &\equiv &<A(x)B(0)>-<A(x)><B(0)> \\
\sigma _{AB}(x) &\equiv &\frac{C_{AB}(x)+C_{BA}(-x)}{2\hbar } \\
\xi _{AB}(x) &\equiv &\frac{C_{AB}(x)-C_{BA}(-x)}{2\hbar }
\end{eqnarray*}
Expectation values are evaluated for free fields in vacuum of quantum field
theory in flat spacetime; symmetrised and antisymmetrised functions are
vacuum expectation values of anticommutator and commutator respectively.
Corresponding spectra are obtained by a translation to momentum domain (same
notation will be used throughout the paper)  
\[
f(x)\equiv \int \frac{{\rm d}^{d}k}{\left( 2\pi \right) ^{d}}
f[k] \exp (-ik_{\mu }x^{\mu }) 
\]
In absence of a more precise specification, spacetime dimension is an
arbitrary integer $d \geq  2$. Vacuum is characterized as the equilibrium
state at zero temperature, with correlation functions obeying
fluctuation-dissipation relations  
\begin{eqnarray}
C_{AB}[k] &=&2\hbar \theta (k_0 )\sigma _{AB}[k]  \nonumber \\
\xi _{AB}[k] &=&\sgn (k_0 )\sigma _{AB}[k]  \eqnum{eq1}
\end{eqnarray}
where $\theta \left( {}\right) $ is Heaviside step function, $\sgn 
\left( {}\right) $ sign function, and $k_0 $ frequency. These relations
mean that vacuum fluctuations do not contain negative frequencies and imply
that, for an atom in vacuum, spontaneous transitions correspond to emission
but not to absorption. Whilst obvious for vacuum field fluctuations \cite
{Qfvst23}, these relations are demonstrated for vacuum stress tensors in a
straightforward manner that we briefly recall. We consider, for simplicity,
the case of a massless field theory where vacuum fluctuations only contain
light-like wavevectors. Stress tensor spectrum is obtained by a convolution
product of field spectra: a wavevector $k$ in the stress tensor spectrum is
a sum ($k=k^\prime +k^{\prime \prime }$) of wavevectors $k^\prime $ and 
$k^{\prime \prime }$ present in a field spectrum. It follows that only
positive frequencies appear in noise spectrum $C_{T_{\mu \nu }T_{\rho \sigma
}}$ of vacuum stress tensor, and also that this spectrum vanishes when $k$
is a spacelike wavevector \cite{Qfvst4}, since it is built from
Lorentz-invariant expressions and contains a factor $\theta (k_0 )$.
Contributions located on the light cone are allowed, and actually arise in
the anomalous case of a two-dimensional spacetime (see appendix A).

Linear response theory has been used to relate dissipative forces
experienced by moving boundaries with force fluctuations felt by motionless
boundaries \cite{Qfvst6}. In the case of coupled metric and stress tensors
considered here, the same formalism allows one to relate the dissipative
response of the stress tensor (or metric tensor) to a perturbation of metric
tensor (or stress tensor) with correlation functions computed in the
unperturbed case. The unperturbed case, which is analogous to the motionless
case for boundaries, corresponds here to free quantum field theories in flat
spacetime. This includes linearized gravitation treated in the spirit of a
field theory \cite{Qfvst12,Qfvst13}, with a metric tensor defined as the sum
of the Minkowski tensor $\eta _{\mu \nu }\equiv {\rm diag}(1,-1,-1,-1)$ and
of a small variation $h_{\mu \nu }$ ($\left| h_{\mu \nu }\right| \ll 1$).
Throughout the paper, we use the Minkowski tensor for raising and lowering 
indices as well as for getting traced tensors.

Linear response theory describes responses to a perturbation, which is here
characterized as a variation $\delta L$ of Lagrangian density  
\begin{equation}
\delta L(x)=-\frac{1}{2}T_{\mu \nu }(x)h^{\mu \nu }(x)  \eqnum{eq2}
\end{equation}
where $T_{\mu \nu }$ is the stress tensor operator. This may be considered
either as a metric perturbation generating a stress tensor response
according to the definition of stress tensor in a metric theory of
gravitation \cite{Qfvst24}, and representing coupling of metric to non
gravitational fields in a linear approximation \cite{Qfvst13}, or as a
stress tensor perturbation generating a metric response according to
Einstein equation  
\begin{equation}
G_{\mu \nu }=\kappa T_{\mu \nu }\qquad \kappa =\frac{8\pi G}{c^2 }
\eqnum{eq3}
\end{equation}
where $G_{\mu \nu }$ is Einstein curvature tensor.

Linear response theory tells us how the susceptibility functions $\chi
_{h_{\mu \nu }h_{\rho \sigma }}$ and $\chi _{T_{\mu \nu }T_{\rho \sigma }}$,
which describe respectively metric and stress tensor responses, are related
to the correlation functions evaluated in flat spacetime. Retarded and
advanced susceptibility functions, conveniently written in momentum domain,
are sums of dispersive and dissipative parts defined as even and odd parts
respectively  
\begin{eqnarray}
\chi _{AB}^{\rm ret}[k] &=&\overline{\chi }_{AB}[k]+i\xi _{AB}[k] 
\nonumber \\
\chi _{AB}^{\rm adv}[k] &=&\chi _{AB}^{\rm ret}[-k]=\overline{\chi }
_{AB}[k]-i\xi _{AB}[k]  \eqnum{eq4}
\end{eqnarray}
Here $A$ and $B$ stand for components of metric or stress tensors.
Dissipative parts $\xi _{AB}$ of these functions are just the commutators
yet appearing in the correlation functions. Response function built
according to Feynman prescription is the retarded susceptibility for
positive frequencies and the advanced one for negative frequencies  
\begin{eqnarray*}
\chi _{AB}[k] &=&\theta (k_0 )\chi _{AB}^{\rm ret}[k]+\theta (-k_0 )\chi
_{AB}^{\rm adv}[k] \\
&=&\overline{\chi }_{AB}[k]+i\sigma _{AB}[k]
\end{eqnarray*}
We have used the fact that $\xi _{AB}$ is an odd function of $k$ and 
$\overline{\chi }_{AB}$ an even one, and used relations (\numeq{1}) for
expressing $\chi _{AB}$ in terms of $\sigma _{AB}$. For the problem of
coupled metric and stress tensors, dissipative functions $\sigma _{AB}$ and 
$\xi _{AB}$ describe particle production in a curved spacetime \cite{Qfvst21}. 
The conditions associated with positiveness of dissipation will be
exhibited later on.

The dispersive function $\overline{\chi }_{AB}$ may in principle be deduced
from the commutator $\xi _{AB}$ through dispersion relations, since
causality implies that the retarded susceptibility is analytic and regular
when frequency lies in the upper half plane $\Im k_0  >0$. In the
case of coupled metric and stress tensors, dispersion relations involve
divergences \cite{Qfvst25} which usually lead to ambiguities in the
extraction of a finite part \cite{Qfvst26}. Dissipative functions are
however unambiguously defined and finite \cite{Qfvst27,Qfvst28}, and we
shall focus our attention on fluctuations and dissipation and disregard
difficulties associated with dispersion relations.

Unavoidable questions are also those of vacuum stability in presence of
gravitation \cite{Qfvst29} and of renormalisability \cite{Qfvst30}.
Presently, a complete and consistent description of quantum gravity is not
available, and its dynamics at high frequencies is poorly understood. In the
present paper, we will restrict our interest to fluctuations and dissipation
at experimentally accessible frequencies, i.e. at frequencies much lower
than Planck scale, and to their effects upon ultimate sensitivity limits. In
this restricted context, we shall see that the well-known properties
discussed in the present section allow one to obtain significant results,
despite of the unsolved problems of quantum gravity. More precisely, we
shall only use Einstein equation (\numeq{3}), which describes effective
gravitation at low frequencies and is directly connected to the form 
(\numeq{2}) of the perturbation, and relations (\numeq{1},\numeq{4}) between
response functions and fluctuations. For properly defined physical
quantities, computation will reveal that there is no ghost in vacuum.

\section*{Proper fluctuations of curvature}

As already stated, Einstein equation can be interpreted as describing metric
response to a stress tensor perturbation and thus used to obtain quantum
fluctuations of metric tensor. In the present section, we write these metric
fluctuations and thereafter deduce curvature fluctuations, which present the
advantage over metric fluctuations to be gauge-independent; notice that, for
gravity, a gauge transformation is a coordinate transformation.

We first write Einstein curvature tensor $G_{\mu \nu }$ in a linear
approximation in $h_{\mu \nu }$ (from now on, quantities like $h_{\mu \nu }$
or $G_{\mu \nu }$ are written in momentum domain)  
\begin{eqnarray*}
G_{\mu \nu } &=&R_{\mu \nu }-\frac{1}2 \eta _{\mu \nu }R\equiv \eta _{\mu
\nu \rho \sigma }R^{\rho \sigma } \\
\eta _{\mu \nu \rho \sigma } &\equiv &\frac{1}2 (\eta _{\mu \rho }\eta
_{\nu \sigma }+\eta _{\mu \sigma }\eta _{\nu \rho }-\eta _{\mu \nu }\eta
_{\rho \sigma }) \\
R_{\mu \nu } &=&\frac{1}2 (k^{2}h_{\mu \nu }+k_{\mu }k_{\nu }h-k_{\mu
}k^{\sigma }h_{\nu \sigma }-k_{\nu }k^{\sigma }h_{\mu \sigma })
\end{eqnarray*}
Here $R_{\mu \nu }$ is linearized Ricci tensor and $h=\eta ^{\mu \nu }h_{\mu
\nu }$.

The linearized solution of Einstein equation (\numeq{3}) may thus be written
in a well chosen gauge, such that $2k^{\mu }h_{\mu \nu }=k_{\nu }h$  
\begin{eqnarray*}
k^{2}h^{\mu \nu } &=&2R^{\mu \nu }=2\overline{\eta }^{\mu \nu \rho \sigma
}\kappa T_{\rho \sigma } \\
\overline{\eta }^{\mu \nu \rho \sigma } &\equiv &\frac{1}2 (\eta ^{\mu \rho
}\eta ^{\nu \sigma }+\eta ^{\mu \sigma }\eta ^{\nu \rho }-\frac{2}{d-2}\eta
^{\mu \nu }\eta ^{\rho \sigma })
\end{eqnarray*}
This solution may be written as the following response equation  
\[
h^{\mu \nu }=\chi _{h^{\mu \nu }h^{\rho \sigma }}T_{\rho \sigma } 
\]
where appears a Feynman response function $\chi _{h^{\mu \nu }h^{\rho \sigma
}}$  
\[
\chi _{h^{\mu \nu }h^{\rho \sigma }}=\frac{2\kappa }{k^2 -i\varepsilon }%
\overline{\eta }^{\mu \nu \rho \sigma } 
\]
Using equations (\numeq{4}), this corresponds to a correlation function $%
\sigma _{h^{\mu \nu }h^{\rho \sigma }}$  
\[
\sigma _{h^{\mu \nu }h^{\rho \sigma }}=2\pi \kappa \delta (k^2 )\overline{%
\eta }^{\mu \nu \rho \sigma } 
\]

With other gauge choices, $\sigma _{h^{\mu \nu }h^{\rho \sigma }}$ may be
written as a sum of a transverse part, which is deduced from the preceding
expression, and of a gauge-dependent longitudinal part (represented by
dots)  
\begin{eqnarray*}
\sigma _{h^{\mu \nu }h^{\rho \sigma }} &=&\pi \kappa \delta (k^2 )(\pi ^{\mu
\rho }\pi ^{\nu \sigma }+\pi ^{\mu \sigma }\pi ^{\nu \rho }-\frac{2}{d-2}\pi
^{\mu \nu }\pi ^{\rho \sigma }) \\
&&+...
\end{eqnarray*}
that is also  
\begin{eqnarray}
\sigma _{h_{\mu \nu }h_{\rho \sigma }} &=&2\pi \kappa \delta (k^2 )\sum
(\lambda _{r}\pi _{\mu \nu \rho \sigma }^{r})+...  \nonumber \\
\lambda _1  &=&-\frac{1}{d-2}\qquad \lambda _0 =1  \eqnum{eq5}
\end{eqnarray}
In these expressions, $\pi _{\mu \nu }$ is the tensorial form of a
Lorentz-invariant transverse tensor  
\[
\pi _{\mu \nu }=\eta _{\mu \nu }-\frac{k_{\mu }k_{\nu }}{k^2 } 
\]
and $\pi _{\mu \nu \rho \sigma }^{r}$ ($r=0,1$) are the two orthogonal
tensorial forms taken by correlation functions of such a tensor  
\begin{eqnarray}
\pi _{\mu \nu \rho \sigma }^{r} &=&\alpha _{r}\pi _{\mu \nu }\pi _{\rho
\sigma }+\beta _{r}(\pi _{\mu \rho }\pi _{\nu \sigma }+\pi _{\mu \sigma }\pi
_{\nu \rho })  \eqnum{eq6} \\
\alpha _1  &=&\frac{1}{d-1}\qquad \beta _1 =0  \nonumber \\
\alpha _0  &=&-\frac{1}{d-1}\qquad \beta _0 =\frac{1}2   \nonumber
\end{eqnarray}
More details about these two forms are given in the next section.

We now evaluate curvature fluctuations. Longitudinal terms do not contribute
to curvature and transverse fluctuations are concentrated upon the light
cone, as for any massless field theory. One checks that, as expected,
fluctuations of Einstein tensor vanish. Curvature fluctuations may therefore
be considered as vacuum gravitational waves \cite{Qfvst14}, characterized
like classical gravitational waves by a non vanishing Riemann curvature and
a vanishing Ricci curvature.

Using the expression of linearized Riemann curvature  
\begin{eqnarray*}
R_{\mu \nu \rho \sigma } &=&\frac{1}2 (k_{\mu }k_{\rho }h_{\nu \sigma
}+k_{\nu }k_{\sigma }h_{\mu \rho }-k_{\nu }k_{\rho }h_{\mu \sigma }-k_{\mu
}k_{\sigma }h_{\nu \rho }) \\
&=&\frac{1}2 (k_{\mu }\eta _{\nu \lambda }-k_{\nu }\eta _{\mu \lambda
})(k_{\rho }\eta _{\sigma \tau }-k_{\sigma }\eta _{\rho \tau })h^{\lambda
\tau }
\end{eqnarray*}
one gets from equations (\numeq{5})  
\begin{eqnarray}
\sigma _{R_{\mu \nu \rho \sigma }R_{\mu ^\prime \nu ^\prime \rho
^\prime \sigma ^\prime }} &=&a{\cal R}_{\mu \nu \rho \sigma }{\cal R}%
_{\mu ^\prime \nu ^\prime \rho ^\prime \sigma ^\prime }  \nonumber \\
&+&b({\cal R}_{\mu \nu \mu ^\prime \nu ^\prime }{\cal R}_{\rho \sigma
\rho ^\prime \sigma ^\prime }+{\cal R}_{\mu \nu \rho ^\prime \sigma
^\prime }{\cal R}_{\rho \sigma \mu ^\prime \nu ^\prime })  \eqnum{eq7}
\end{eqnarray}
Coefficients ${\cal R}_{\mu \nu \rho \sigma }$ correspond to Riemann
curvatures evaluated for a conformal metric ($h_{\mu \nu }=\eta _{\mu \nu }$%
)  
\begin{eqnarray*}
{\cal R}_{\mu \nu \rho \sigma } &=&\frac{1}2 (k_{\mu }\eta _{\nu \lambda
}-k_{\nu }\eta _{\mu \lambda })(k_{\rho }\eta _{\sigma \tau }-k_{\sigma
}\eta _{\rho \tau })\eta ^{\lambda \tau } \\
&=&\frac{1}2 (k_{\mu }k_{\rho }\eta _{\nu \sigma }+k_{\nu }k_{\sigma }\eta
_{\mu \rho }-k_{\nu }k_{\rho }\eta _{\mu \sigma }-k_{\mu }k_{\sigma }\eta
_{\nu \rho })
\end{eqnarray*}
Coefficients $a$ and $b$ are  
\begin{eqnarray}
a &=&2\pi \kappa \delta (k^2 )\sum (\lambda _{r}\alpha _{r})=-\frac{2}{d-2}%
\pi \kappa \delta (k^2 )  \nonumber \\
b &=&2\pi \kappa \delta (k^2 )\sum (\lambda _{r}\beta _{r})=\pi \kappa
\delta (k^2 )  \eqnum{eq8}
\end{eqnarray}
One may consider these equations as a Lorentz-invariant and
gauge-independent form of already known expressions for vacuum gravitational
waves (see for example metric fluctuations evaluated in transverse traceless
gauge in \cite{Qfvst14}).

\section*{Fluctuations and dissipative polarization of vacuum stress tensors}

We now study the correlation functions characterizing quantum fluctuations
of vacuum stress tensors of non gravitational fields. We then express the
dissipative response of vacuum stress tensor to a metric perturbation in
terms of these correlation functions.

The correlation functions of vacuum stress tensors are evaluated in
Minkowski spacetime according to the general prescription of linear response
theory (see above). Their tensorial properties follow from symmetries of
stress tensor, Lorentz invariance and conservation laws for vacuum fields in
Minkowski spacetime. As a consequence of the latter property $T_{\mu \nu }$
is divergenceless, i.e. transverse in momentum domain, and correlation
functions obey  
\begin{equation}
k^{\mu }\sigma _{T_{\mu \nu }T_{\rho \sigma }}=k^{\rho }\sigma _{T_{\mu \nu
}T_{\rho \sigma }}=0  \eqnum{eq9}
\end{equation}
It follows \cite{Qfvst21} that spectra decompose over the two transverse
tensorial forms $\pi _{r}$ defined by equations (\numeq{6}) 

\begin{eqnarray}
\sigma _{T_{\mu \nu }T_{\rho \sigma }} &=&(k^2 )^2 \sigma \sum (\zeta
_{r}\pi _{\mu \nu \rho \sigma }^{r})  \nonumber \\
\sigma &=&\hbar \pi (k^2 )^{d/2-2}\theta (k^2 )  \eqnum{eq10}
\end{eqnarray}
Note however that an exception arises for $d=2$. Extra terms are indeed
allowed for wavevectors localized on the light cone, and such terms
effectively contribute for scalar fields in two-dimensional spacetime (see
appendix A).

The two tensorial forms $\pi ^{r}$ are orthogonal projectors onto the
subspace of transverse tensors. They obey simple rules for product and trace
operations  
\begin{eqnarray*}
(\pi ^{r}\cdot \pi ^{s})_{\mu \nu \rho \sigma } &\equiv &\pi _{\mu \nu
}^{r}{}^{\lambda \tau }\pi _{\lambda \tau \rho \sigma }^{s}=\delta _{rs}\pi
_{\mu \nu \rho \sigma }^{r} \\
\left( \pi ^1 \right) _{\mu \rho \sigma }^{\mu } &=&\pi _{\rho \sigma
}\qquad \left( \pi ^0 \right) _{\mu \rho \sigma }^{\mu }=0
\end{eqnarray*}
Applying these two projectors onto the stress tensor, one therefore gets a
decomposition of its fluctuations as a sum of two uncorrelated components  
\begin{eqnarray*}
T_{\mu \nu } &=&T_{\mu \nu }^1 +T_{\mu \nu }^0  \\
T_{\mu \nu }^1  &=&(\pi ^1 \cdot T)_{\mu \nu }=\pi _{\mu \nu }^1 {}^{\rho
\sigma }T_{\rho \sigma }=\frac{1}{d-1}\pi _{\mu \nu }T \\
T_{\mu \nu }^0  &=&(\pi ^0 \cdot T)_{\mu \nu }=\pi _{\mu \nu }^0 {}^{\rho
\sigma }T_{\rho \sigma }=T_{\mu \nu }-\frac{1}{d-1}\pi _{\mu \nu }T
\end{eqnarray*}
Component $T_{\mu \nu }^1 $ is proportional to the traced tensor ($T\equiv
\eta ^{\rho \sigma }T_{\rho \sigma }$) while component $T_{\mu \nu }^0 $ is
traceless. Contributions to noise of these two components correspond to the
two parts ($r=0,1$) in equation (\numeq{10}).

The two coefficients $\zeta _{r}$ ($r=0,1$) appearing in equations 
(\numeq{10}) are dimensionless functions of $k^2 $, and pure numbers for
massless field theories. Note that we use natural spacetime units with $c=1$%
, but we keep $\hbar $ as a scale for quantum fluctuations in equations 
(\numeq{10}) and similar forthcoming expressions. Explicit expressions of these
coefficients computed for scalar fields and Maxwell fields in a spacetime of
arbitrary dimension are given in appendix A. In the following, we shall
consider $\zeta _{r}$ as the sum of contributions of all non gravitational
fields, restricting our interest to massless fields which correspond to a
long range polarization. We will touch on gravitational contribution later
on.

Now, the dissipative part of the stress tensor response to a metric
perturbation is described by  
\[
<\delta T_{\mu \nu }>_{{\rm dissip}}=\sigma _{T_{\mu \nu }T_{\rho \sigma
}}h^{\rho \sigma }\equiv (\sigma _{TT}\cdot h)_{\mu \nu } 
\]
This is the dissipative part of Feynman response function; for retarded and
advanced responses, $\sigma _{TT}$ has to be replaced by $\xi _{TT}$ (for
the sake of readibility, we occasionally omit indices). As a consequence of
relations (\numeq{9}), this expression vanishes for metric variations
corresponding to coordinate transformations  
\[
h^{\rho \sigma }=-i(k^{\rho }\xi ^{\sigma }+k^{\sigma }\xi ^{\rho })\quad
\rightarrow \quad <\delta T_{\mu \nu }>_{{\rm dissip}}=0 
\]
In other words, dissipative part of vacuum stress tensor polarization has an
intrinsically geometrical (i.e. gauge-independent) character, which appears
closely connected to energy conservation for vacuum fluctuations in
Minkowski spacetime. It is not affected by coordinate transformations and
may thus be written in terms of curvatures only.

Using equation (\numeq{10}), we may obtain explicit expressions of vacuum
stress tensor polarization in terms of spacetime curvatures, more precisely
of the two independent transverse tensors $G_{\mu \nu }^{r}$ that can be
built from Einstein tensor ($G^{r}\equiv \pi ^{r}\cdot G$; $\lambda _{r}$
defined in equations \numeq{5})  
\begin{equation}
<\delta T_{\mu \nu }>_{{\rm dissip}}=2k^2 \sigma \sum (\lambda _{r}\zeta
_{r}G_{\mu \nu }^{r})  \eqnum{eq11}
\end{equation}
Component $G_{\mu \nu }^1 $ is proportional to traced Einstein curvature ($%
G=\eta ^{\mu \nu }G_{\mu \nu }$), or equivalently to scalar curvature ($%
R=\eta ^{\mu \nu }R_{\mu \nu }$)  
\[
G_{\mu \nu }^1 =\frac{1}{d-1}\pi _{\mu \nu }G\qquad G=-\frac{d-2}{2}R 
\]
In contrast, $G_{\mu \nu }^0 $ is traceless and can be written in terms of
Weyl curvature  
\[
G_{\mu \nu }^0 =G_{\mu \nu }-\frac{1}{d-1}\pi _{\mu \nu }G=\frac{d-2}{d-3}%
\frac{k^{\rho }k^{\sigma }}{k^2 }W_{\mu \rho \nu \sigma } 
\]
Weyl tensor vanishes for $d=2$ or $d=3$ and is otherwise defined by  
\begin{eqnarray*}
W_{\mu \rho \nu \sigma } &=&R_{\mu \rho \nu \sigma } \\
&-&\frac{1}{d-2}(\eta _{\rho \sigma }R_{\mu \nu }+\eta _{\mu \nu }R_{\rho
\sigma }-\eta _{\rho \nu }R_{\mu \sigma }-\eta _{\mu \sigma }R_{\rho \nu })
\\
&+&\frac{1}{d-1}\frac{1}{d-2}(\eta _{\mu \nu }\eta _{\rho \sigma }-\eta
_{\mu \sigma }\eta _{\rho \nu })R
\end{eqnarray*}
These two components thus have different behaviours with respect to
conformal metric perturbations ($h_{\mu \nu }\sim \eta _{\mu \nu }$), since
only $G_{\mu \nu }^1 $ differs from zero in this case. Note also that only $%
G_{\mu \nu }^0 $ contributes to stress tensor polarization for conformally
invariant field theories (see appendix A).

It may be emphasized that vacuum stress tensor polarization (\numeq{11})
cannot be written as a local expression \cite{Qfvst28}, because of factor $%
\theta (k^2 )$ appearing in expression (\numeq{10}) of $\sigma $. Using
fluctuation-dissipation relations (\numeq{1}) and translating from momentum
domain to spacetime domain, one shows that it has rather a causally
propagating form with retarded and advanced responses propagating on the
light cone. In fact, this factor $\theta (k^2 )$ may be considered as
expressing causality of vacuum response to a metric perturbation.

As discussed in the introduction, dissipative vacuum stress tensor
polarization may be identified with particle production in response to a
gravitational perturbation. It is thus expected to obey some positiveness
conditions which imply that perturbation is damped by back-reaction of
particle production. In order to exhibit such conditions, it is worth
studying energy-momentum transfer between matter fields and gravity \cite
{Qfvst21}. As well known, one can define a pseudo stress tensor, say $\Theta
_{\mu \nu }$, for gravitational field, which is such that the total stress
tensor $T_{\mu \nu }$ + $\Theta _{\mu \nu }$ obeys an ordinary
energy-momentum conservation law \cite{Qfvst31}. Since the matter stress
tensor $T_{\mu \nu }$ has a null covariant divergence, one obtains the
energy-momentum transfer as  
\begin{eqnarray*}
\partial ^{\nu }T_{\mu \nu }(x) &=&-\partial ^{\nu }\Theta _{\mu \nu }(x) \\
&=&\Gamma _{\nu }^{\lambda \nu }(x)T_{\mu \lambda }(x)+\Gamma _{\mu
}^{\lambda \nu }(x)T_{\nu \lambda }(x)
\end{eqnarray*}
where $\Gamma _{\mu }^{\lambda \nu }$ ($=\eta ^{\nu \rho }\Gamma _{\mu \rho
}^{\lambda }$) are Christoffel symbols. This transfer vanishes at first
order in the metric tensor, because the linearized stress tensor has a null
divergence and $\Theta _{\mu \nu }$ is a quadratic expression of the metric
tensor. It may be evaluated at second order by replacing on the right-hand
side the Christoffel symbols and stress tensors by their linearized
expressions. One thus obtains the energy-momentum transfer $\Pi _{\mu }$
integrated over spacetime as  
\begin{eqnarray}
\Pi _{\mu } &=&\int \frac{{\rm d}^{d}k}{\left( 2\pi \right) ^{d}}\hbar
k_{\mu } \sgn (k_0 )n[k]  \nonumber \\
n[k] &=&\frac{1}{2\hbar }\sigma _{T_{\lambda \nu }T_{\rho \sigma
}}[k]h^{\rho \sigma }[k]h^{\lambda \nu }[-k]  \eqnum{eq12}
\end{eqnarray}
This integrated transfer $\Pi _{\mu }$ characterizes energy-momentum
dissipation due to particle production in response to a metric perturbation; 
$\hbar k_{\mu }$ is the energy-momentum transfer per produced particle while 
$n[k]$ is the density of particles produced at a given wavevector. Due to
transversality (\numeq{9}) of stress tensor correlation functions, function $%
n[k]$ may be written as a scalar quadratic form of curvatures  
\[
n[k]=2\pi (k^2 )^{d/2-2}\theta (k^2 )\sum (\zeta _{r}\lambda _{r}^{2}G^{r\
\mu \nu }[k]G_{\mu \nu }^{r}[-k]) 
\]
with  
\begin{eqnarray*}
4G^{1\ \mu \nu }G_{\mu \nu }^1  &=&\frac{(d-2)^2 }{d-1}R^2  \\
4G^{0\ \mu \nu }G_{\mu \nu }^0  &=&\frac{d-2}{d-3}W^{\mu \rho \nu \sigma
}W_{\mu \rho \nu \sigma } \\
&=&R^{\mu \rho \nu \sigma }R_{\mu \rho \nu \sigma }-\frac{R^2 }{d-1}
\end{eqnarray*}
Positiveness of dissipated energy-momentum thus appears to be associated
with positiveness of the coefficients $\zeta _{r}$. One effectively checks
that coefficients $\zeta _{r}$ given in appendix A are positive.

\section*{Quantum fluctuations of coupled metric and stress tensors}

Stress tensor and spacetime curvature are coupled dynamical systems. Up to
now, we have studied the proper fluctuations of each system, as well as the
response of a system to a perturbation due to the other one. The purpose of
the present section is to give a consistent description at lowest orders of
coupled fluctuations and of interlinked response mechanisms. Stress tensor
responds to curvatures, i.e. to transverse components of the metric tensor.
We may therefore restrict the discussion to transverse parts of the
correlation functions and deal with the two orthogonal components ($r=0,1$)
separately.

We first write coupled stress tensor fluctuations as sums of corresponding
input fluctuations ($T_{\mu \nu }^{r\ {\rm in}}$) and of linear response to metric
fluctuations  
\begin{equation}
T_{\mu \nu }^{r}=T_{\mu \nu }^{r\ {\rm in}}+\chi _{TT}^{r\ {\rm in}}h_{\mu
\nu }^{r}  \eqnum{13a} 
\end{equation}
Input fluctuations are characterized by correlation functions (\numeq{10})
where the subscript ``in'' did not appear  
\[
\sigma _{T_{\mu \nu }T_{\rho \sigma }}{}^{{\rm in}}=\sum (\sigma _{TT}^{r\ 
{\rm in}}\pi _{\mu \nu \rho \sigma }^{r})\qquad \sigma _{TT}^{r\ {\rm in}%
}=(k^2 )^2 \zeta _{r}\sigma 
\]
Linear susceptibility $\chi _{TT}^{r\ {\rm in}}$ is the lowest order vacuum
stress tensor polarization  
\[
\chi _{TT}^{r\ {\rm in}}=(k^2 )^2 \zeta _{r}\Gamma _{r}=(k^2 )^2 \zeta
_{r}(\overline{\Gamma }_{r}+i\sigma ) 
\]
Correlation function $\sigma _{TT}^{r\ {\rm in}}$ is the imaginary part of
the linear susceptibility $\chi _{TT}^{r\ {\rm in}}$. As already discussed,
we will not make use of the detailed form of the dispersive part $\overline{%
\Gamma }_{r}$.

In a similar manner, we write the linear response of metric to stress tensor
fluctuations  
\begin{equation}
h_{\mu \nu }^{r}=h_{\mu \nu }^{r\ {\rm in}}+\chi _{hh}^{r\ {\rm in}}T_{\mu
\nu }^{r}  \eqnum{13b}  
\end{equation}
Input fluctuations $h_{\mu \nu }^{r\ {\rm in}}$ are characterized by a
correlation function $\sigma _{hh}^{r\ {\rm in}}$ which is the imaginary
part of the linear susceptibility $\chi _{hh}^{r\ {\rm in}}$ (longitudinal
terms are omitted)  
\begin{eqnarray*}
\sigma _{h_{\mu \nu }h_{\rho \sigma }}{}^{{\rm in}} &=&\sum (\sigma
_{hh}^{r\ {\rm in}}\pi _{\mu \nu \rho \sigma }^{r})\qquad \sigma _{hh}^{r\ 
{\rm in}}=2\pi \kappa \lambda _{r}\delta (k^2 ) \\
\chi _{hh}^{r\ {\rm in}} &=&\frac{2\kappa \lambda _{r}}{k^2 -i\varepsilon }=%
\frac{2\kappa \lambda _{r}}{k^2 }-1+i\sigma _{hh}^{r\ {\rm in}}
\end{eqnarray*}

One easily solves equations (\numeq{13}) to obtain coupled
fluctuations in terms of input ones  
\begin{eqnarray*}
T_{\mu \nu }^{r} &=&\chi _{Th}^{r}T_{\mu \nu }^{r\ {\rm in}}+\chi
_{TT}^{r}h_{\mu \nu }^{r\ {\rm in}} \\
h_{\mu \nu }^{r} &=&\chi _{Th}^{r}h_{\mu \nu }^{r\ {\rm in}}+\chi
_{hh}^{r}T_{\mu \nu }^{r\ {\rm in}}
\end{eqnarray*}
with  
\begin{eqnarray*}
\chi _{hh}^{r} &=&\frac{1}{(\chi _{hh}^{r\ {\rm in}})^{-1}-\chi _{TT}^{r\ 
{\rm in}}} \\
\chi _{Th}^{r} &=&\frac{1}{1-\chi _{TT}^{r\ {\rm in}}\chi _{hh}^{r\ {\rm in}}%
} \\
\chi _{TT}^{r} &=&\frac{1}{(\chi _{TT}^{r\ {\rm in}})^{-1}-\chi _{hh}^{r\ 
{\rm in}}}
\end{eqnarray*}
One deduces the correlation functions for coupled fluctuations  
\begin{eqnarray*}
&&\sigma _{h_{\mu \nu }h_{\rho \sigma }}=\sum (\sigma _{hh}^{r}\pi _{\mu \nu
\rho \sigma }^{r}) \\
\sigma _{hh}^{r} &=&\chi _{hh}^{r}\sigma _{TT}^{r\ {\rm in}}(\chi
_{hh}^{r})^{*}+\chi _{Th}^{r}\sigma _{hh}^{r\ {\rm in}}(\chi _{Th}^{r})^{*}
\end{eqnarray*}
Fluctuation-dissipation relations for coupled variables follow from those
known for input ones: coupled correlation function $\sigma _{hh}^{r}$ is the
imaginary part of coupled susceptibility $\chi _{hh}^{r}$, and coupled
correlation functions obey relations (\numeq{1}) characteristic of vacuum.
Similar results are obtained for stress tensor correlation functions and
cross correlations between stress tensor and metric.

Writing coupled susceptibilities as  
\begin{eqnarray*}
\chi _{hh}^{r} &=&\gamma _{r}\chi _{hh}^{r\ {\rm in}}\qquad \chi
_{TT}^{r}=\gamma _{r}\chi _{TT}^{r\ {\rm in}} \\
\chi _{Th}^{r} &=&\gamma _{r}=\frac{1}{1-2\kappa \lambda _{r}\zeta
_{r}k^2 \Gamma _{r}}
\end{eqnarray*}
one sees that $\kappa \gamma _{r}$ is a momentum-dependent effective
gravitational constant for the $r-$component, which may be compared with the
effective mass for an unbound mirror coupled to vacuum radiation pressure 
\cite{Qfvst10}. Equal low frequency values $\gamma _1 [0]$ and $\gamma
_0 [0] $, differing from the standard value 1, could be dealt with by
redefining $\kappa $. A difference between $\gamma _1 [0]$ and $\gamma
_0 [0]$ would lead to an effective gravitation differing from the
predictions of general relativity. This possibility is discussed in appendix
B. Since accurate experiments have checked that gravitation is consistent
with Einstein theory \cite{Qfvst32}, we will assume in the following that  
\[
\gamma _1 [0]=\gamma _0 [0]=1 
\]

We remark that coupled metric fluctuations may be written  
\[
\sigma _{hh}^{r}=\sigma _{hh}^{r\ {\rm in}}+\left( \frac{2\kappa \lambda _{r}%
}{k^2 }\right) ^2 \sigma _{TT}^{r} 
\]
The former term coincides with proper metric fluctuations while the latter
appears to result from gravity of vacuum stress tensor. As expected, these
two contributions have been included in a consistent treatment. At this
point, it is worth recalling that metric fluctuations have a non-commutative
character as vacuum stress tensor fluctuations (see equations \numeq{1}),
and noting that metric and stress tensor fluctuations are correlated in the
coupled system.

We now restrict the discussion to momenta much lower than Planck frequency,
where correlation functions have simple approximated forms  
\begin{eqnarray*}
\sigma _{h_{\mu \nu }h_{\rho \sigma }} &=&\sigma _{h_{\mu \nu }h_{\rho
\sigma }}{}^{{\rm in}}+4\kappa ^2 \sigma \sum (\lambda _{r}^2 \zeta
_{r}\pi _{\mu \nu \rho \sigma }^{r}) \\
\sigma _{T_{\mu \nu }T_{\rho \sigma }} &=&\sigma _{T_{\mu \nu }T_{\rho
\sigma }}{}^{{\rm in}}=(k^2 )^2 \sigma \sum (\zeta _{r}\pi _{\mu \nu \rho
\sigma }^{r}) \\
\sigma _{T_{\mu \nu }h_{\rho \sigma }} &=&2\kappa k^2 \sigma \sum (\lambda
_{r}\zeta _{r}\pi _{\mu \nu \rho \sigma }^{r})
\end{eqnarray*}
We deduce correlations of Einstein curvature and stress tensor  
\begin{eqnarray*}
\sigma _{G_{\mu \nu }G_{\rho \sigma }} &=&\kappa ^2 \sigma _{T_{\mu \nu
}T_{\rho \sigma }} \\
\sigma _{G_{\mu \nu }T_{\rho \sigma }} &=&\kappa \sigma _{T_{\mu \nu
}T_{\rho \sigma }}
\end{eqnarray*}
which can be summarized by simple identities for stochastic variables  
\[
G_{\mu \nu }=\kappa T_{\mu \nu }=\kappa T_{\mu \nu }{}^{{\rm in}} 
\]
We eventually obtain correlation functions for Riemann curvature
fluctuations. Their form is still given by equations (\numeq{7}) with
coefficients $a$ and $b$ being sums of proper terms (\numeq{8}) and of terms
arising from gravity of vacuum stress tensor  
\begin{eqnarray}
a &=&-\frac{2}{d-2}\pi \kappa \delta (k^2 )+4\kappa ^2 \sigma \sum
(\lambda _{r}^2 \zeta _{r}\alpha _{r})  \nonumber \\
b &=&\pi \kappa \delta (k^2 )+4\kappa ^2 \sigma \sum (\lambda
_{r}^2 \zeta _{r}\beta _{r})  \eqnum{14}
\end{eqnarray}

In these expressions, $\zeta _{r}$ is a sum of contributions of non
gravitational fields, which we may restrict for simplicity to massless
fields. The gravitational contribution deserves a specific treatment. The
pseudo stress tensor $\Theta _{\mu \nu }$ for gravitational field is a
quadratic expression of the metric tensor, but it has the same magnitude as
the stress tensor $T_{\mu \nu }$ associated with non gravitational fields,
and must in principle be taken into account in the analysis of vacuum stress
tensor fluctuations and associated polarization. At first sight, it seems
natural to conclude that coefficients $\zeta _{r}$ have to be modified in
order to include a gravitational contribution \cite{Qfvst33}. The non linear
nature of gravitation however entails a second correction to the previously
computed expressions. Without entering into a detailed discussion of these
corrections, we want to emphasize the following point. Since $\Theta _{\mu
\nu }$ is directly related to the non linear correction to Einstein tensor,
the already written expressions are a correct description of the
fluctuations of Einstein and Ricci curvatures in terms of non gravitational
stress tensors only.

\section*{Quantum limits in spacetime probing}

In this final section, we analyse in detail how curvature fluctuations
result in sensitivity limits in spacetime probing.

For that purpose, we consider the specific case of four-dimensional
spacetime and rewrite Riemann curvature fluctuations  
\begin{eqnarray*}
C_{R_{\mu \nu \rho \sigma }R_{\mu ^\prime \nu ^\prime \rho ^{\prime
}\sigma ^\prime }} &=&4\beta ({\cal R}_{\mu \nu \mu ^\prime \nu ^{\prime
}}{\cal R}_{\rho \sigma \rho ^\prime \sigma ^\prime }+{\cal R}_{\mu \nu
\rho ^\prime \sigma ^\prime }{\cal R}_{\rho \sigma \mu ^\prime \nu
^\prime }) \\
&-&4\alpha {\cal R}_{\mu \nu \rho \sigma }{\cal R}_{\mu ^\prime \nu
^\prime \rho ^\prime \sigma ^\prime }
\end{eqnarray*}
where functions $\alpha $ and $\beta $\ are obtained by using equations 
(\numeq{1}), (\numeq{7}), (\numeq{14}) and definitions (\numeq{5}), (\numeq{6})
and (\numeq{10}), and by considering only massless fields, i.e. the Maxwell
field and $N_{\nu }$ massless neutrino fields which each contribute for one
fourth of the contribution of Maxwell field \cite{Qfvst28}  
\begin{eqnarray*}
\alpha &=&4\pi ^{2}l_{P}^2 \theta (k_0 )
\left( \delta (k^2 )+\frac{4+N_{\nu }}{%
30\pi }l_{P}^2 \theta (k^2 ) \right) \\
\beta \ &=&4\pi ^{2}l_{P}^2 \theta (k_0 )
\left( \delta (k^2 )+\frac{4+N_{\nu }}{%
20\pi }l_{P}^2 \theta (k^2 ) \right) 
\end{eqnarray*}
More precisely, $N_{\nu }$ has to be understood as the number of neutrino
fields with a mass smaller than noise frequencies (see appendix A); note
also that $\zeta _1 $ vanishes for massless fields, as a consequence of
conformal invariance. Noise spectra $C_{R_{\mu \nu \rho \sigma }R_{\mu
^\prime \nu ^\prime \rho ^\prime \sigma ^\prime }}$ contain
components proportional to $\delta (k^2 )$ which correspond to vacuum
gravitational waves and scale as $(kl_{P})^2 $, and components proportional
to $\theta (k^2 )$ which arise from gravity of vacuum stress tensors and
scale as $(kl_{P})^{4}$. The latter have a smaller magnitude at low momenta,
but are present in a larger momentum domain.

In length or time measurements using a probe field, for example in
interferometric or timing measurements, the field phase registers metric
perturbations along propagation. Using the law of geodesic deviation \cite
{Qfvst15}, this effect is described by a deviation tensor $\Delta _{\mu \rho
}$ in the eikonal approximation and at first order in curvature  
\begin{eqnarray*}
\Delta _{\mu \rho } &\equiv &\frac{1}{K_0 }\frac{\partial K_{\rho }}{%
\partial x^{\mu }}\equiv \frac{1}{K_0 }\frac{\partial K_{\mu }}{\partial
x^{\rho }} \\
 &=&\int_0 ^{\tau }Q_{\mu \rho }\left( x-t\frac{K}{K_0 } \right) 
{\rm d}t \\
Q_{\mu \rho } &=&R_{\mu \nu \rho \sigma }\frac{K^{\nu }K^{\sigma }}{K_0 ^2 %
}
\end{eqnarray*}
The integral is evaluated along a one-way track (other measurement
techniques are discussed in \cite{Qfvst16,Qfvst34} and references therein),
the coordinate time $t$ is used as an affine parameter and $\tau $ is the
time of propagation from the emitter to the receiver. $K_{\rho }$ is the
wavevector of the probe field and $K_0 $ its frequency; $K_{\rho }$ is
related to the gradient of the probe phase, or equivalently to the
four-dimensional velocity vector for the probe beam. Notice that the
deviation tensor $\Delta _{\mu \rho }$ and the tidal tensor $Q_{\mu \rho }$
are homogeneous functions of $K$, in the eikonal approximation. The effect
of curvature fluctuations upon length or time measurements may then be
characterized by noise spectra for components of the deviation tensor  
\begin{eqnarray*}
C_{\Delta _{\mu \nu }\Delta _{\rho \sigma }}[\omega ] &=&
\left( \frac{\tau }{2\pi } \right)^2 
\int \frac{{\rm d}{\bf k}^2 {\rm d}k_{3}}2  \\
\times &<&C_{Q_{\mu \nu }Q_{\rho \sigma }}>_{\omega ,{\bf k}^2 ,k_{3}}{\rm %
sinc}^2 \frac{(\omega -k_{3}v)\tau }2 
\end{eqnarray*}
Deviation tensors are evaluated at a given spatial position. For simplicity,
we consider from now on that the probe propagates along the $x_{3}-$axis
with a normalized velocity $v$ ($K_1 =K_2 =0$; $K_{3}=K_{0}v$; $v=1$ for a
massless probe field; $v<1$ for a massive probe field) and identify the
various components as temporal (index 0), longitudinal (index 3) and
transverse (index 1 or 2). In the foregoing equation, ${\rm sinc}(x)$ stands
for $\frac{\sin (x)}{x}$ and $<C_{Q_{\mu \nu }Q_{\rho \sigma }}>_{\omega ,%
{\bf k}^2 ,k_{3}}$ represents the average of $C_{Q_{\mu \nu }Q_{\rho \sigma
}}[k]$ over azimut angle with the constraints $k_0 =\omega $ and $%
k_1 ^2 +k_2 ^2 +k_{3}^2 ={\bf k}^2 $.

Tidal tensor $Q_{\mu \rho }$ is obtained through a contraction of the
Riemann tensor defined with respect to the propagation direction of the
probe. Temporal components of the deviation tensor may be expressed in terms
of longitudinal ones  
\begin{eqnarray*}
Q_{03} &=&vQ_{33}\qquad Q_{00}=v^{2}Q_{33} \\
Q_{01} &=&vQ_{13}\qquad Q_{02}=vQ_{23}
\end{eqnarray*}
As a consequence, temporal components vanish at the limit of slow test
particles, where the effect of curvature is reduced to a purely spatial
tidal effect. More generally for any velocity, it will be sufficient to
study the 6 spatial components of the tidal tensor, whose noise spectra are
given by  
\begin{eqnarray*}
\sigma _{Q_{ij}Q_{kl}} &=&4\beta ({\cal Q}_{ik}{\cal Q}_{jl}+{\cal Q}_{il}%
{\cal Q}_{jk})-4\alpha {\cal Q}_{ij}{\cal Q}_{kl} \\
{\cal Q}_{ik} &=&\frac{1}2 \left( k_{i}k_{k}(1-v^2 )-\delta
_{ik}(k_0 -k_{3}v)^2 \right. \\
&&\left. -v(\delta _{i3}k_{k}+k_{i}\delta _{k3})(k_0 -k_{3}v)\right)
\end{eqnarray*}
Latin indices represent spatial components and $\delta _{ik}$ is the
Kronecker symbol for such indices. In place of $Q_{11}$ and $Q_{22}$, we
introduce the two variables  
\begin{eqnarray*}
&&Q_{12}^\prime =\frac{Q_{11}-Q_{22}}2  \\
Q &=&Q_{\mu }^{\mu }=-(Q_{11}+Q_{22})-Q_{33}(1-v^2 ) \\
&=&R_{\nu \sigma }\frac{K^{\nu }K^{\sigma }}{K_0 ^2 }
\end{eqnarray*}

Straightforward computations lead to noise spectra characterizing
fluctuations of the components of the tidal tensor averaged over azimut
angle. It turns out that the 4 components $Q_{13}$, $Q_{23}$, $Q_{12}$ and $%
Q_{12}^\prime $ are uncorrelated stochastic variables with noise spectra
given by  
\begin{eqnarray*}
<C_{Q_{13}Q_{13}}> &=&<C_{Q_{23}Q_{23}}> \\
&=&\frac{\beta -\alpha }2 ({\bf k}^2 -k_{3}^2 )(\omega -k_{3}v)^2  \\
&-&\frac{2\beta -\alpha }2 ({\bf k}^2 -k_{3}^2 )(\omega
^2 -k_{3}^2 )(1-v^2 ) \\
&+&\beta (\omega ^2 -k_{3}^2 )(\omega -k_{3}v)^2  \\
<C_{Q_{12}Q_{12}}> &=&<C_{Q_{12}^{\prime }Q_{12}^\prime }> \\
&=&\frac{2\beta -\alpha }{8}({\bf k}^2 -k_{3}^2 )^2 (1-v^2 )^2  \\
&+&\beta (\omega -k_{3}v)^{4} \\
&-&\beta (\omega -k_{3}v)^2 ({\bf k}^2 -k_{3}^2 )(1-v^2 )
\end{eqnarray*}
The 2 remaining components are correlated variables with noise spectra given
by  
\begin{eqnarray*}
<C_{Q_{33}Q_{33}}> &=&(2\beta -\alpha )(\omega ^2 -k_{3}^2 )^2  \\
<C_{Q_{33}Q}> &=&-(2\beta -\alpha )(\omega ^2 -{\bf k}^2 )(\omega
^2 -k_{3}^2 )(1-v^2 ) \\
&-&2(\beta -\alpha )({\bf k}^2 -k_{3}^2 )(\omega -k_{3}v)^2  \\
&+&2\alpha (\omega ^2 -{\bf k}^2 )(\omega -k_{3}v)^2  \\
<C_{QQ}> &=&(2\beta -\alpha )(\omega ^2 -{\bf k}^2 )^2 (1-v^2 )^2  \\
&+&4(\beta -\alpha )(\omega -k_{3}v)^{4} \\
&-&4\alpha (\omega ^2 -{\bf k}^2 )(1-v^2 )(\omega -k_{3}v)^2 
\end{eqnarray*}

\section*{Discussion}

Propagation of a massless field is resonantly affected by fluctuations with
light-like wavevectors, so that the smaller fluctuations with time-like
wavevectors may be disregarded in this case. For a massive probe field in
contrast, propagation is not resonantly affected by curvature fluctuations.

If we restrict our attention to the contribution of vacuum gravitational
waves (light-like wavevectors), the foregoing expressions are simplified due
to the relations $\beta =\alpha $; $\omega ^2 ={\bf k}^2 $. Component $Q$
thus vanishes 
\[
<C_{Q_{33}Q}>=<C_{QQ}>=0
\]
since Ricci curvature vanishes for
gravitational waves. There remain 5 uncorrelated components with spectra
given by  
\begin{eqnarray*}
<C_{Q_{33}Q_{33}}> &=&\alpha (\omega ^2 -k_{3}^2 )^2  \\
<C_{Q_{13}Q_{13}}> &=&<C_{Q_{23}Q_{23}}> \\
&=&\alpha (\omega ^2 -k_{3}^2 )(\omega -k_{3}v)^2  \\
&-&\frac{\alpha }2 (\omega ^2 -k_{3}^2 )^2 (1-v^2 ) \\
<C_{Q_{12}Q_{12}}> &=&<C_{Q_{12}^{\prime }Q_{12}^\prime }> \\
&=&\alpha (\omega -k_{3}v)^{4} \\
&-&\alpha (\omega ^2 -k_{3}^2 )(1-v^2 )(\omega -k_{3}v)^2  \\
&+&\frac{\alpha }{8}(\omega ^2 -k_{3}^2 )^2 (1-v^2 )^2  \\
\alpha &=&4\pi ^{2}l_{P}^2 \theta (\omega )\delta (\omega ^2 -{\bf k}^2 )
\end{eqnarray*}
Component $Q_{33}$ describes a longitudinal deviation for longitudinally
shifted geodesics, which plays a role in interferometric measurements. This
is why its noise spectrum has already been studied in the particular case of
massless probes ($v=1$), as a way to the detection of a stochastic
background of gravitational waves \cite{Qfvst34}, and in the context of
ultimate quantum limits in optical interferometry or in electromagnetic
timing experiments \cite{Qfvst16}. Other components of the tidal and
deviation tensors have apparently not been studied up to now. Components $%
Q_{13}$ and $Q_{23}$ correspond to transverse deviations for longitudinally
shifted geodesics, or equivalently to longitudinal deviations for
transversely shifted geodesics, that is to a bending of the probe beam.
Components $Q_{12} $ and $Q_{12}^\prime $ correspond to transverse
deviations for transversely shifted geodesics, that is to a focusing or
defocusing effect of the probe beam.

We may emphasize that the foregoing expressions also contain informations
about ultimate quantum limits in measurements based upon atomic
interferometry ($v<1$). In the limit of slow probe particles ($v\rightarrow
0 $) in particular, the geodesic deviation tensor takes a simple form, and
the leading contributions to noise spectra due to vacuum gravitational waves
may be written  
\begin{eqnarray*}
C_{\Delta _{33}\Delta _{33}}[\omega ] &=&\frac{32}{15}\omega
^{3}l_{P}^2 \sin ^2 \frac{\omega \tau }2 \theta (\omega ) \\
C_{\Delta _{13}\Delta _{13}}[\omega ] &=&C_{\Delta _{23}\Delta _{23}}[\omega
]=C_{\Delta _{12}\Delta _{12}}[\omega ] \\
&=&C_{\Delta _{12}^\prime \Delta _{12}^\prime }[\omega ]=\frac{3}{4}%
C_{\Delta _{33}\Delta _{33}}[\omega ]
\end{eqnarray*}

When compared to foregoing expressions, higher order curvature fluctuations
due to gravity of vacuum correspond to small modifications scaling as $%
\omega ^{5}l_{P}^{4}$. A new feature associated with these higher order
terms is that fluctuations of the trace component ($\Delta =\Delta _{\mu
}^{\mu }$) no longer vanish  
\[
C_{\Delta \Delta }[\omega ]=\frac{8(4+N_{\nu })}{105\pi }\omega
^{5}l_{P}^{4}\sin ^2 \frac{\omega \tau }2 \theta (\omega ) 
\]
Notice that this noise spectrum would not be affected by the contribution of
gravitational stress tensor, since $\Delta $ is obtained as a contraction of
Ricci tensor (see the discussion in the end of section 5). It is worth
emphasizing that this expression depends on the number $N_{\nu }$ of
massless neutrino fields, precisely of neutrino fields with a mass smaller
than the analysis frequency.

In the context of this paper where attention is restricted to fluctuations
and dissipation at experimentally accessible frequencies, this last result
establishes a direct connection between fundamental matter fields and
minimal fluctuations of spacetime. This connection has been derived here
from a few minimal properties of gravitational quantum fluctuations, namely
the relations between response functions and fluctuations and the known
effective behaviour of gravitation at low frequencies. It should therefore
subsist in a consistent theory including quantum gravity.

\appendix 

\section{Stress tensor correlation functions for a spacetime of arbitrary
dimension}

Stress tensor correlation functions can be written for a spacetime of
arbitrary integer dimension $d$ ($d\geq  2$). A purely dimensional effect
appears in expression (\numeq{10}) of $\sigma $. Coefficients $\zeta _{r}$
also depend upon the dimension and the specific quantum field theory studied.

Massless scalar fields correspond to  
\begin{eqnarray*}
\zeta _0  &=&(4\pi )^{-d/2}\frac{\Gamma (\frac{d}2 +1)}{\Gamma (d+2)} \\
\zeta _1  &=&\frac{(d-2)^2 (d+1)}2 \zeta _0 
\end{eqnarray*}
The particular case of a two-dimensional spacetime requires a special
attention. Explicit computation gives  
\[
\sigma _{T_{\mu \nu }T_{\rho \sigma }}=\frac{\hbar }{24}\pi _{\mu \nu \rho
\sigma }^{0}k^2 \theta (k^2 )+\frac{\hbar }{12}k_{\mu }k_{\nu }k_{\rho
}k_{\sigma }\delta (k^2 ) 
\]
This expression obeys condition (\numeq{9}) associated with energy
conservation. In contrast with the generic case however, an extra
contribution describing fluctuations with light-like wavevectors appears
superimposed to the $\pi ^{r}-$terms ($\zeta _1 =0$ and $\zeta _0 =\frac{1%
}{24\pi }$\ in this case). This reveals the anomalous character of results
specific to a two-dimensional spacetime.

It is possible to compute the functions $\zeta _{r}$ for massive fields
(mass $\mu $; $\zeta _{r}\{\mu =0\}$ is given above)  
\begin{eqnarray*}
\zeta _{r} &=&\zeta _{r}\{\mu =0\}\theta (k^2 -4\mu ^2 ) \\
&&\times \left( 1-4\frac{\mu ^2 }{k^2 }\right) ^{(d-3)/2}\left( 1-4\lambda
_{r}\frac{\mu ^2 }{k^2 }\right) ^2 
\end{eqnarray*}
Note that positive values are obtained for any values of momenta.

One may also give stress tensor correlation functions for Maxwell fields  
\begin{eqnarray*}
\zeta _0  &=&(4\pi )^{-d/2}\frac{\Gamma (\frac{d}2 +1)}{\Gamma (d+2)}%
(2d^2 -3d-8) \\
\zeta _1  &=&\frac{(4\pi )^{-d/2}}2 \frac{\Gamma (\frac{d}2 +1)}{\Gamma
(d+2)}(d-4)^2 (d-2)(d+1)
\end{eqnarray*}
For two-dimensional spacetime, $\zeta _0 $ is negative, in contrast with
the generic positiveness property of coefficients $\zeta _{r}$. In this case
however, Weyl curvatures vanish and do not appear in stress tensor response.

We get in particular for massless scalar fields in four-dimensional
spacetime (same result as in \cite{Qfvst28})  
\[
\zeta _0 =\frac{1}{960\pi ^2 }\qquad \zeta _1 =10\zeta _0  
\]
and for Maxwell fields in four-dimensional spacetime (same result as in \cite
{Qfvst27,Qfvst28})  
\[
\zeta _0 =\frac{1}{80\pi ^2 }\qquad \zeta _1 =0 
\]
Coefficient $\zeta _1 $ vanishes in this latter case, as a consequence of
conformal invariance of Maxwell equations. It also vanishes for massless
scalar fields in a two-dimensional spacetime, for the same reason.

\section{Effective gravitation and vacuum stress tensor polarization}

Effective gravitation theory may be modified by vacuum stress tensor
polarization, if $\gamma _1 $(0) and g0(0) differ from the standard value
1. Some results of the paper have to be changed in this case. The graviton
propagator is obtained as  
\[
\chi _{h_{\mu \nu }h_{\rho \sigma }}=\frac{2\kappa }{k^2 -i\varepsilon }%
\sum (\lambda _{r}\gamma _{r}\pi _{\mu \nu \rho \sigma }^{r})+\ldots 
\]
and corresponds to a modified gravitation equation  
\[
G_{\mu \nu }^{r}=\kappa \gamma _{r}T_{\mu \nu }^{r} 
\]
Defining an effective gravitational constant  
\[
\kappa _{\rm eff}=\kappa \gamma _0 (0) 
\]
and a parameter measuring deviation from general relativity  
\[
\delta \gamma _1 =\frac{\gamma _1 (0)-\gamma _0 (0)}{\gamma _0 (0)} 
\]
we rewrite a modified Einstein equation  
\[
G_{\mu \nu }^{r}=\kappa _{\rm eff}(T_{\mu \nu }+\delta \gamma _{1}T_{\mu
\nu }^1 ) 
\]
As $T_{\mu \nu }^1 $ is proportional to the traced stress tensor, this
equation looks like a scalar-tensor equation of gravitation \cite{Qfvst32}.

For evaluating the deviation from general relativity, we study the metric
perturbation in the field of a static point-like mass $m$  
\begin{eqnarray*}
h_{00}[k] &=&2\pi \delta (k_0 )\frac{\kappa _{\rm eff}m}{k^2 }\frac{2}{%
d-2}\left( d-3-\frac{\delta \gamma _1 }{d-1}\right) \\
h_{ii}[k] &=&2\pi \delta (k_0 )\frac{\kappa _{\rm eff}m}{k^2 }\frac{2}{%
d-2}\left( 1+\frac{\delta \gamma _1 }{d-1}\right)
\end{eqnarray*}
Latin indices represent spatial components and metric tensor is evaluated at
first order in $m$ with a gauge choice such that it is diagonal. We then
compute Eddington parameter $\gamma $ as the ratio between the spatial and
temporal perturbations, and obtain for $d=4$  
\[
\gamma =\frac{1+\frac{\delta \gamma _1 }{3}}{1-\frac{\delta \gamma _1 }{3}}
\]
Experiments on light bending and gravitational time delays \cite{Qfvst32}
tell us that $\delta \gamma _1 $ has a small value and then that $\gamma
_1 (0)$ and $\gamma _0 (0)$ are close to each other.

Correlation functions for proper fluctuations of Riemann curvatures are
given by equations (\numeq{7}) with modified coefficients $a$ and $b$  
\begin{eqnarray*}
a &=&2\pi \kappa \delta (k^2 )\sum (\lambda _{r}\gamma _{r}\alpha _{r}) \\
&=&-\frac{2}{d-2}\pi \kappa _{\rm eff}\delta (k^2 )\left( 1+\frac{\delta
\gamma _1 }{d-1}\right) \\
b &=&2\pi \kappa \delta (k^2 )\sum (\lambda _{r}\gamma _{r}\beta _{r}) \\
&=&\pi \kappa _{\rm eff}\delta (k^2 )
\end{eqnarray*}
Einstein tensor does no longer vanish but the curvature tensor $\sum \frac{%
G_{\mu \nu }^{r}}{\gamma _{r}}$ is directly proportional to the stress
tensor and therefore has vanishing fluctuations.

\endmc

\end{document}